\documentstyle[12pt]{article}

\begin{document}
\title{\textbf{Ricci dynamo stretch-shear plasma flows and magnetic energy bounds}} \maketitle
{\sl \textbf{L.C. Garcia de Andrade}

Departamento de F\'{\i}sica
Te\'orica-IF-Universidade do Estado do Rio de Janeiro\\[-3mm]
Rua S\~ao Francisco Xavier, 524\\[-3mm]
Cep 20550-003, Maracan\~a, Rio de Janeiro, RJ, Brasil\\[-3mm]
Electronic mail address: garcia@dft.if.uerj.br\\[-3mm]
\vspace{0.1cm} \newline{\bf Abstract} \paragraph*{
   Geometrical tools, used in Einstein's general relativity (GR), are applied to dynamo theory, in order to obtain fast dynamo action bounds to magnetic energy, from Killing symmetries in Ricci flows. Magnetic field is shown to be the shear flow tensor eigendirection, in the case of marginal dynamos. Killing symmetries of the Riemann metric, bounded by Einstein space, allows us to reduce the computations. Techniques used are similar to those strain decomposition of the flow in Sobolev space, recently used by Nu\~nez [JMP \textbf{43} (2002)] to place bounds in the magnetic energy in the case of hydromagnetic dynamos with plasma resistivity. Contrary to Nu\~nez case, we assume that the dynamos are kinematic, and the velocity flow gradient is decomposed into expansion, shear and twist. The effective twist vanishes by considering that the frame vorticity coincides with Ricci rotation coefficients. Eigenvalues are here Lyapunov exponents. In analogy to GR, where curvature plays the role of gravity, here Ricci curvature seems to play the role of diffusion.}
\newpage
\section{Introduction}
In analogy to Chiconne, Latushkin and Montgomery-Schmidt \cite{1}, on the fast dynamo operator in Riemannian manifolds, a new fast dynamo operator is defined on Ricci flows. The magnetic field growth rate is computed in terms of eigenvalues of the Ricci tensor in Einstein spaces \cite{2}. Dynamo fast action is obtained when the real part of the magnetic growth rate is positive. Recently, M Nu\~nez \cite{3} has developed a method of obtaining dynamo action in Sobolev spaces by considering that the using a number of functional inequalities, to estimate the rate of increase of magnetic energy in terms of plasma resistivity and different norms of plasma velocity strain. In this paper techniques of Einstein gravity (GR) as the ones recently applied to vortex filaments by Ricca \cite{4}, called Ricci rotation coefficients and Killing symmetries of the metric tensor, are used to obtain the a Ricci flow, which helps us to obtain the appropriated bounds to the magnetic energy as eigendirections of shear tensor of the flow. Recently Thiffeault and Boozer \cite{5,6} have shown that eigendirections are fundamental in the investigation of chaotic and advection processes in the kinematic dynamos in Riemann manifolds. Actually the ideas of simple dynamo
models in curved space were inspired by the cat
Arnold's chaotic dynamo
first toy model \cite{7} uniformly stretched in Riemannian space. 
Riemann metric of the flow does not need to be necessarily diagonal, and may be of that used by Chui and Moffatt \cite{8} of the knotted magnetic flux tubes. Other covariant model, important in magnetic reconnection of astrophysical plasmas has been recently developed by Titov et al \cite{9}. More recently \cite{10}, one has investigated the existence of slow dynamos in the diffusive plasma Riemannian space, with resistivity. In the present paper diffusion vanishes, and the chaotic dynamos are considered. When the expansion or contraction of the flow, called stretching in dynamo theory, vanishes, the curvature sign plays a fundamental role in the existence of dynamo action as has been shown by Chiconne at al \cite{1}, where Riemannian Anosov manifolds of constant negative curvature. Bayly and Childress \cite{11} has also consider the importance of the shear for dynamo action in Euclidean space. The paper is organized as follows: Section II discusses the fast dynamo operator in Ricci flows. In Section III the analysis of the Killing symmetries and bounds in the magnetic energy, is undertaken. Section IV presents future prospects and conclusions.
\section{Fast dynamo operator in Ricci flows}
This section presents the Riemann metric given by the Ricci flow \cite{12} given in mathematical terms by\newline
\textbf{Definition II.1}:
\begin{equation}
\frac{{\partial}\textbf{g}}{{\partial}t}=-2\textbf{Ric}
\label{1}
\end{equation}
where here, $\textbf{g}$ is the Riemann metric, over manifold $\cal{M}$, and the parameter t in $\textbf{g}(t)$, is given in the interval $t\in{[a,b]}$ in the field of real numbers $\textbf{R}$. On a local chart $\cal{U}$ in $\cal{M}$, the expression (\ref{1}) can be expressed as \cite{15}
\begin{equation}
\frac{{\partial}{g_{ij}}}{{\partial}t}=-2{R_{ij}}
\label{2}
\end{equation}
where $\textbf{Ric}$, is the Ricci tensor, whose components $R_{ij}$. From this expression, one defines the eigenvalue problem as
\begin{equation}
R_{ij}{\chi}^{j}={\lambda}{\chi}_{i}\label{3}
\end{equation}
where $(i,j=1,2,3)$. Substitution of the Ricci flow equation (\ref{2}) into this eigenvalue expression and cancelling the eigendirection ${\chi}^{i}$ on both sides of the equation yields
\begin{equation}
\frac{{\partial}g_{ij}}{{\partial}t}=-2\lambda{g_{ij}}
\label{4}
\end{equation}
Solution of this equation yields the Lyapunov expression for the metric
\begin{equation}
g_{ij}=exp{[-2\lambda{t}]}{\delta}_{ij}
\label{5}
\end{equation}
where ${\delta}_{ij}$ is the Kroenecker delta. Note that in principle if ${\lambda}\le{0}$ the metric grows without bounds, and in case it is negative it is bounded as $t\rightarrow{\infty}$. Recently Thiffeault has used a similar Lyapunov exponents expression in Riemannian manifolds to investigate chaotic flows, without attention to dynamos or Ricci flow. Thus one has proven the following lemma:\newline
\textbf{Lemma II.1}:\newline
If ${\lambda}_{i}$ is an eigenvalue spectra of the $\textbf{Ric}$ tensor, the finite-time Lyapunov exponents spectra is given by
\begin{equation}
{\lambda}_{i}=-{\gamma}_{i}\le{0}\label{6}
\end{equation}
In the remaining of the paper the Lyapunov eigenvalues shall play an important role in the determination of the bounds of magnetic energy as a global dynamo action bound. As an important step in this direction, let us consider the chaotic magnetic kinematic dynamo with zero plasma resistivity
\begin{equation}
\frac{d\textbf{B}}{dt}=\textbf{B}.{\nabla}\textbf{v}\label{7}
\end{equation}
where $\textbf{B}$ is the magnetic field vector, and $\textbf{v}$ is the flow velocity. In general curvilinear coordinates ${x^{i}}\in{\cal{U}}_{i}$ in the subchart $U_{i}$ of the manifold, in the rotating frame reference of the flow $\textbf{e}_{i}$, this equation may be written as
\begin{equation}
\frac{d(B^{i}\textbf{e}_{i})}{dt}={B^{i}}{\partial}_{i}({v^{j}\textbf{e}_{j}})\label{8}
\end{equation}
where 
\begin{equation}
\textbf{B}=B^{i}\textbf{e}_{i}\label{9}
\end{equation}
From the evolution of the reference frame 
\begin{equation}
\frac{d\textbf{e}_{i}}{dt}={{\omega}_{i}}^{j}\textbf{e}_{j}\label{10}
\end{equation}
and the Ricci rotation coefficient
\begin{equation}
{{\partial}_{k}}\textbf{e}_{i}={{\Gamma}_{ki}}^{j}\textbf{e}_{j}\label{11}
\end{equation}
where ${\omega}_{ij}$ and ${{\Gamma}_{ki}}^{j}$ are respectively the vorticity and the Ricci rotation coefficients, one is able to expand expression (\ref{8}) as 
\begin{equation}
\frac{dB^{i}}{dt}={B^{p}}[{\partial}_{p}(g^{il}){v_{l}}+g^{il}({\nabla}_{p}v_{l}-{\omega}_{pl})]
\label{12}
\end{equation}
where
\begin{equation}
{\nabla}_{p}v_{l}={\partial}_{p}v_{l}+{{\Gamma}_{pl}}^{k}v_{k}
\label{13}
\end{equation}
is the covariant derivative of the Ricci flow component. As usual in cosmological fluids, we decompose the covariant derivative of the flow in the invariant decomposition of vorticity ${\Omega}_{lp}$, shear ${\sigma}_{kl}$ tensors and expansion ${\theta}$
as
\begin{equation}
{\nabla}_{p}v_{l}={\Omega}_{pl}+{\sigma}_{pl}-\frac{1}{3}{\theta}g_{lk}
\label{14}
\end{equation}
Substitution of this expression into the self-induction equation (\ref{12}) yields
\begin{equation}
\frac{dB^{i}}{dt}={B^{p}}[{\partial}_{p}(g^{il}){v_{l}}+g^{il}[({\Omega}_{pl}-{\omega}_{pl})+{\sigma}_{pl}-\frac{1}{3}{\theta}g_{lk}]]
\label{15}
\end{equation}
This equation can be simplified if one assumes that the flow has a rigid rotation, or that the vorticity of the flow coincides with the vorticity of the frame or
\begin{equation}
{\Omega}_{pl}={\omega}_{pl}
\label{16}
\end{equation}
This choice reduces the self-induction equation to
\begin{equation}
\frac{dB^{i}}{dt}={B^{p}}[{\partial}_{p}(g^{il}){v_{l}}+g^{il}[{\sigma}_{pl}-\frac{1}{3}{\theta}g_{lp}]]
\label{17}
\end{equation}
This expression allows us to obtain the growth rate ${\gamma}$ defined as
\begin{equation}
B^{i}={B^{0}}exp[{\gamma}t+i{\omega}_{k}x^{k}]
\label{18}
\end{equation}
After substituting expression (\ref{18}) into (\ref{17}), a simple algebra yields
\begin{equation}
{\gamma}=[{\sigma}+{\Omega}-\frac{1}{3}{\theta}]
\label{19}
\end{equation}
Note that vorticity, or twist, still survives. Here ${\sigma}$, ${\Omega}$ and ${\theta}$ are the shear (fold), vorticity (twist) \cite{13} and expansion or compression (stretching) eigenvalues of the respective tensors and scalar as
\begin{equation}
{\sigma}_{ki}B^{i}={\sigma}B_{k}
\label{20}
\end{equation}
\begin{equation}
{\Omega}_{ki}B^{i}={\Omega}B_{k}
\label{21}
\end{equation}
\begin{equation}
{\Theta}_{ki}B^{i}=-\frac{1}{3}{\theta}B_{k}
\label{22}
\end{equation}
Thus one notices that the fast operator can be defined from these expressions as:
\newline
\textbf{Lemma II.2}: 
\newline
The fast dynamo operator of the Ricci flow is given by
\begin{equation}
{\cal{L}}=\frac{d}{dt}-[{\sigma}-\frac{1}{3}{\theta}+{\omega}]
\label{23}
\end{equation}
Note that expression (\ref{19}) shows that twist and shear is enhanced by the stretching ${\theta}\le{0}$. The Ricci flow eigenvalue is not present in the last expression, however in the next section, one shall observe that in the computation of the magnetic energy the Ricci flow eigenvalue shall be present and contribute to bound the magnetic energy and therefore, the dynamo action.
\section{Ricci flows bounds to dynamo action}
In this section one shall use the last self-induction equation, to compute the magnetic energy ${\epsilon}$ as
\begin{equation}
{\epsilon}=\int{B^{2}dV}
\label{24}
\end{equation}
which expressed in terms of the 3D Riemann metric components reads
\begin{equation}
{\epsilon}=\int{B^{i}g_{ij}B^{j}dV}
\label{25}
\end{equation}
Since, by definition fast dynamo action corresponds to the growth of magnetic energy in time as $\frac{{\partial}{\epsilon}}{{\partial}t}\ge{0}$ , this amount has to be computed by performing the partial time derivative of the expression (\ref{25}). Actually the equal sign in the last condition represents the lower limit of marginal dynamos, where the magnetic energy integral remains constant. This computation yields
\begin{equation}
\frac{{\partial}{\epsilon}}{{\partial}t}=\frac{{\partial}[\int{B^{i}g_{ij}B^{j}dV}]}{{\partial}t}
\label{26}
\end{equation}
Expansion of the RHS of this expression shows clearly now where the Ricci flow eigenvalue effect is going to appear. This long computation, reduces the energy integral to
\begin{equation}
\frac{{\partial}{\epsilon}}{{\partial}t}=\int{[2{\sigma}_{pl}+\frac{1}{2}
({\Omega}_{lp}-\frac{1}{3}g_{lp}{\theta})-4R_{lp}]B^{p}B^{l}dV}
\label{27}
\end{equation}
The appearence of the Ricci tensor came exactly from the time derivative of the metric, while the self-induction equation of the last section was also used. In this computation the following Killing symmetries ${\partial}_{l}g^{l3}=0$ and the flow constraint $v^{l}={{\delta}^{l}}_{3}$, in this steady flow. By considering, now that the Ricci tensor is constrained to an Einstein space where
\begin{equation}
R_{lp}={\Lambda}g_{lp}
\label{28}
\end{equation}
substitution of this expression yields
\begin{equation}
\frac{{\partial}{\epsilon}}{{\partial}t}=\int{[2{\sigma}_{pl}+\frac{1}{2}
({\Omega}-(\frac{1}{3}{\theta})+4{\Lambda}))g_{lp}]B^{p}B^{l}dV}
\label{29}
\end{equation}
Note therefore that the vanishing of the integrand yields a marginal dynamo
\begin{equation}
{\sigma}_{pl}=-\frac{1}{4}[{\Omega}-\frac{1}{3}{\theta}+4{\Lambda})]g_{lp}
\label{30}
\end{equation}
Thus note that even in the absence of stretching and twist, shear would still be present by the action of Ricci flow. This seems to be a new kind of effect not present in previous flows. The expression for the shear tensor in this case reduces to
\begin{equation}
{\sigma}_{pl}=-{\Lambda}g_{lp}
\label{31}
\end{equation}
Thus if the space is Einstein-flat, ${\Lambda}$ vanishes and so is the shear. In general fast dynamo action could be obtained from the inequality
\begin{equation}
{\sigma}+\frac{1}{4}[{\Omega}-\frac{1}{3}{\theta}+4{\Lambda}]\ge{0}\label{32}
\end{equation}
Thus even in the case of unstretching and vorticity, shear in Ricci flows seems to be enough to support dynamo action.
\section{Conclusions}
 By making use of mathematical tools from Riemannian geometry, popular in Einstein general relativity, called Ricci Rotation Coefficients, and Killing symmetries, one obtains fast dynamo action in unstretching magnetic field lines, which seems to contradict Vishik's \cite{14} anti-fast dynamos which was recently shown to be held also in resistivity plasmas. This aspect deserves however, further detailed investigation. Magnetic helicity, in Ricci flows manifolds may also appear someplace else. Dynamic effects on the stretching of the magnetic field by a plasma flow, has been recently investigated by Nu\~nez \cite{15} in diffusion media. It is interesting to observe here that the curvature introduced by the Ricci flow, seems to play the role of a diffusion media in analogy to what happens in GR. Another interesting aspect of the model discussed here, is that the Lyapunov structure of the metric is conformal, and conformal geometrical dynamos can be also found in the literature \cite{16}. Another interesting aspect of future work maybe the investigation of anti-dynamo theorems in Ricci plasma flows \cite{17}.
\section{Acknowledgements}
I appreciate financial  supports from UERJ and CNPq.


\newpage

\begin{thebibliography}{17}
  \bibitem{1} C. Chicone and Yu Latushkin, Evolution Semigroups in Dynamical systems and differential equations, American
  Mathematical Society, AMS-(1999). C. Chicone and Yu Latushkin and S. Montgomery-Smith,Comm. Math. Physics \textbf{173} 379 (1995). C. Chicone and Yu Latushkin, Proc of the American
  Mathematical Society 125, N. 11,3391 (1997).
  \bibitem{2} A Busse, Einstein manifolds, (1980) Springer-Verlag. 
  \bibitem{3} M Nu\~nez, J Mathematical Physics \textbf{43}.
  \bibitem{4} R Ricca, Phys Rev \textbf{A 43}, 4281 (1991) .
  \bibitem{5} J L Thiffeault and A H Boozer, Chaos \textbf{11} 16 (2001).
  \bibitem{6} J L Thiffeault, Differential constraints in Chaotic Flows on Curved Manifolds, (2002) arXiv:nlin/0209042.v1.
  \bibitem{7} V. Arnold and B. Khesin, Topological Methods in Hydrodynamics,
  Applied Mathematics Sciences 125 (1991) Springer. V. Arnold, Ya B. Zeldovich, A. Ruzmaikin and D.D.
  Sokoloff, JETP 81 (1981),n. 6, 2052. V. Arnold, Ya B. Zeldovich, A. Ruzmaikin and D.D.
  Sokoloff, Doklady Akad. Nauka SSSR 266 (1982) n6, 1357.
  \bibitem{8} Y Chui and K H Moffatt, Proc Roy Soc London \textbf{A 451} 609 (1995).
  \bibitem{9} V Titov, T Forbes, E R Priest, Z Mikic and J Linder, ApJ \textbf{631}, 1029 (2009). V Titov, Ap J \textbf{660},863 (2007).
  \bibitem{10} L.C. Garcia de Andrade, Phys Plasmas \textbf{15},(2008) and 14, 102902
  (2007).
  \bibitem{11} S. Childress and A. D. Gilbert,\textbf{Stretch, Twist, Fold:
  The fast dynamo}, Springer, Berlin, New York, (1995).
  \bibitem{12} B Chow and D Knopf, Ricci Flows:An introduction, (2004) AMS, New York.
  \bibitem{13} S. I. Vainshtein, Ya B Zeldovich, Sov Phys Usp 15
  ,159 (1972).
  \bibitem{14} M. Vishik, Izv Acad Science,USSR Phys Solid Earth \textbf{24} 173 (1988).
  \bibitem{15} M Nu\~nez, J Phys \textbf{A},8903 (2003). 
  \bibitem{16} L.C. Garcia de Andrade, Phys. Plasmas \textbf{14},(2007). L C Garcia de Andrade, Phys Plasmas \textbf{13}
  (2006).
  \bibitem{17} L.C. Garcia de Andrade, The role of stretching and
  curvature in fast dynamo plasmas in Riemannian space, Phys Plasmas \textbf{15} (2008) in press. 
  \end{thebibliography}
  \end{document}